\documentclass[twocolumn,prl,aps,showpacs,floatfix]{revtex4}

\usepackage{graphicx}
\usepackage{revsymb}
\usepackage{dcolumn}

\begin{document}

\title{Comparative study of depth-dose distributions for beams of light and heavy nuclei in tissue-like media}

\author{Igor~Pshenichnov\footnotemark\footnotetext{Corresponding author: pshenich@fias.uni-frankfurt.de}}
   \affiliation{Frankfurt Institute for Advanced Studies, J.-W. Goethe University, 
                60438 Frankfurt am Main, Germany }
   \affiliation{ Institute for Nuclear Research, Russian Academy of Science, 117312 Moscow, Russia}
\author{Igor~Mishustin}
   \affiliation{Frankfurt Institute for Advanced Studies, J.-W. Goethe University, 
                60438 Frankfurt am Main, Germany }
   \affiliation{Kurchatov Institute, Russian Research Center, 123182 Moscow, Russia}
\author{Walter~Greiner}
   \affiliation{Frankfurt Institute for Advanced Studies, J.-W. Goethe University, 
                60438 Frankfurt am Main, Germany }
\date{\today}

\begin{abstract}
We study the energy deposition by light and heavy nuclei in tissue-like media as used for cancer
therapy. The depth-dose distributions for protons, $^{3}$He, $^{12}$C, $^{20}$Ne, and $^{58}$Ni nuclei are
calculated within a Monte Carlo model based on the GEANT4 toolkit.
These distributions are compared with each other and with available experimental data. 
It is demonstrated that nuclear fragmentation reactions essentially reduce the peak-to-plateau ratio of 
the dose profiles for deeply penetrating energetic ions heavier than $^{3}$He.
On the other hand, all projectiles up to $^{20}$Ne were found 
equally suitable for therapeutic use at low penetration depths.
\end{abstract}


\maketitle

Proton and ion beams of intermediate energies are widely used nowadays for 
cancer treatment~\cite{Amaldi:Kraft:2005}.
Irradiation of deeply-seated tumors without destroying healthy tissues becomes
possible because such heavy projectiles deliver enhanced dose at the very end of their 
range in tissues, close to the Bragg peak. In the first pioneering studies, 
see Ref.~\cite{Chen:Castro:Quivey:1981} for a historical review, 
the choice of projectile nuclei and beam energy for radiation therapy
was mainly determined by the parameters of accelerators available in physics laboratories.
Several experimental studies with homogeneous phantoms
were also performed. Accurate measurements of the depth-dose distributions for $^{12}$C, 
$^{18}$O and $^{20}$Ne ions in water were made at GSI, Germany and at NIRS, 
Japan~\cite{Sihver:Schardt:Kanai:1998}. 

It is expected~\cite{Amaldi:Kraft:2005} that carbon ions are more advantageous in radiotherapy 
compared to protons because of (1) reduced longitudinal and lateral scattering in tissues, 
(2) increased relative biological effectiveness (RBE) close to the Bragg peak, and 
(3) the possibility to monitor the beam range by the positron emission tomography (PET).
The latter is realized via the detection of positron emission by nuclear fragments 
(e.g. $^{10}$C and $^{11}$C) created in fragmentation of beam nuclei.
Currently GSI is using  $^{12}$C beams for treatment and
continues research with such beams in different phantoms~\cite{Haettner:Iwase:Schardt:2006}.
Two new dedicated hospitals in Heidelberg, Germany, and in Pavia, Italy, will soon 
provide cancer treatments with carbon and proton beams~\cite{Amaldi:Kraft:2005}. 
Two hospital-based facilities in Japan, HIMAC at Chiba and HIBMC at Hyogo are using 
carbon beams for cancer treatment too~\cite{Amaldi:Kraft:2005}. 
The feasibility of $^{3}$He beams
for therapy is currently under investigation~\cite{Fiedler:etal:2006}.     
      
The aim of this paper is to provide theoretical guidance in choosing the ion kind and beam energy 
from the view point of their suitability for cancer therapy. 
This is done by comparison of depth-dose distributions for various projectiles, from
protons to nickel ions at different energies. We take into
account electromagnetic and hadronic interactions of primary and secondary particles.
We pay special attention to nuclear fragmentation reactions which reduce the fluence of primary beam ions.  
We believe that the lack of systematic experimental studies with ions other than carbon can be 
partially filled with calculations.


The propagation of protons and nuclei in tissue-like media is studied with a Monte 
Carlo Model for Heavy-ion Therapy (MCHIT) based on the GEANT4 
toolkit~\cite{Agostinelli:etal:2003,Allison:etal:2006} (version 8.2). 
Here we briefly describe the choice of models employed in our calculations, as more details
can be found in Refs.~\cite{Pshenichnov:etal:2005,Pshenichnov:etal:2006}. 

The energy loss of primary and secondary charged particles due to electromagnetic 
interactions is calculated with a set of Monte Carlo models called 'standard electromagnetic physics'. 
This accounts for energy loss and straggling of charged particles due to interaction with atomic 
electrons as well as multiple Coulomb scattering on atomic nuclei.   
At each simulation step, the ionisation energy loss of a charged particle is calculated according to the
Bethe-Bloch formula. There the mean excitation potential of 
water molecules was set to 77 eV, i.e. to the value which better describes the set of available data on 
depth-dose distributions measured with carbon-ion 
beams~\cite{Sihver:Schardt:Kanai:1998,Haettner:Iwase:Schardt:2006}.

The binary cascade model~\cite{Agostinelli:etal:2003,Allison:etal:2006} is used to describe 
the collisions of energetic nucleons and ions with protons and nuclei of the medium. 
After the cascade stage of interaction the decay of excited 
nuclear remnants is considered by employing several models. 
The Weisskopf-Ewing model is used for the description of evaporation of nucleons from nuclei at
excitation energies below 3 MeV per nucleon. The Statistical Multifragmentation Model 
(SMM)~\cite{Bondorf:etal:1995} is used to describe multi-fragment break-up of 
highly-excited residual nuclei at excitation energies above 3 MeV per nucleon.
The SMM includes as its part the Fermi break-up model for 
describing an explosive disintegration of highly-excited light nuclei 
with $Z\le 8$ and $A\le 18$.  

\begin{figure}[htb]  
\begin{centering}
\includegraphics[width=1.\columnwidth]{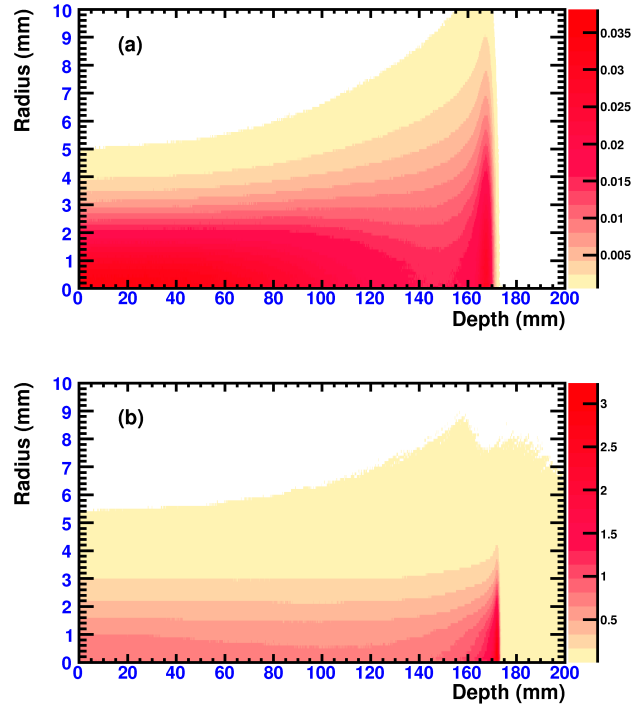}
\caption{Color online. MCHIT simulation of volume dose distributions, in MeV/mm$^3$ per beam particle,   
for 4~mm FWHM beams of (a) 170 MeV protons and (b) 330 A MeV $^{12}$C ions in PMMA.
Notice the dose scales are different in two panels.  
 }
\label{fig:radial}
\end{centering}
\end{figure}

Below we present some results obtained with the MCHIT model.
Figure~\ref{fig:radial} shows 3D dose distributions for the proton and 
$^{12}$C beams with Gaussian intensity profiles of
4~mm FWHM propagating through polymethylmethacrylate (PMMA). 
The dose is defined as the deposited energy per unit volume per beam projectile 
and measured in MeV per mm$^3$.
The beam energies are chosen so that the stopping point of both projectiles 
is located at the same distance of $\sim 170$ mm, corresponding
to the Bragg peak in linear energy deposition. 
Secondary nucleons and nuclear fragments are produced at different depths and 
have broad angular and energy distributions. Their effect must be carefully evaluated, especially 
in the regions beyond the Bragg peak. One can clearly see that multiple scattering of beam particles 
leads to radial widening of the dose field. This effect is considerably larger for
protons than for carbon ions.
The MCHIT model provides the possibility for accurate calculations of the spatial
dose distributions as needed for cancer therapy. 

\begin{figure}[htb]  
\begin{centering}
\includegraphics[width=1.1\columnwidth]{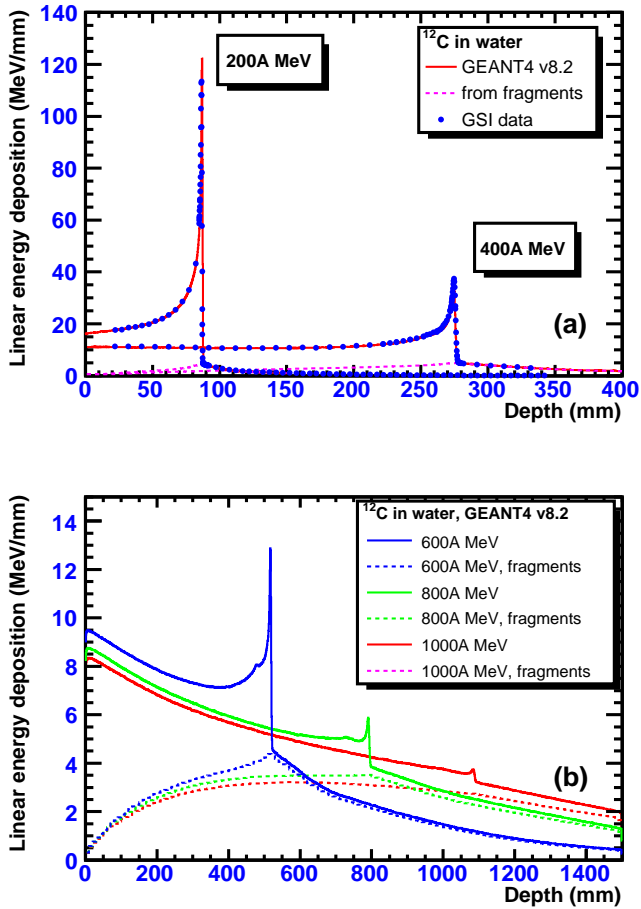}
\caption{Color online. Calculated depth-dose distributions for (a) 200 and 400 A MeV and (b) 
600, 800 and 1000 A MeV $^{12}$C 
beams in water (solid histograms). Contributions from all charged fragments
with $Z\neq6$ are shown by dashed histograms.   
Corresponding experimental data~\cite{Haettner:Iwase:Schardt:2006} are shown by points.}
\label{fig:C12_ENERGY_TWO_PANEL}
\end{centering}
\end{figure}

As carbon-ion beams are currently widely used in heavy-ion therapy,
it is important to know well the evolution of their depth-dose profiles with increasing beam energy.
Corresponding distributions for 200 and 400 A MeV  $^{12}$C nuclei 
calculated with the MCHIT model are shown in 
Fig.~\ref{fig:C12_ENERGY_TWO_PANEL}(a) and compared to the experimental 
data~\cite{Haettner:Iwase:Schardt:2006}. Results for
600, 800 and 1000 A MeV  $^{12}$C nuclei are shown in Fig.~\ref{fig:C12_ENERGY_TWO_PANEL}(b). 
The depth-dose distributions were obtained in terms of the average linear energy 
deposition per beam particle and expressed in MeV/mm.  They were calculated  
by splitting a cubic phantom into thin slices and calculating the 
energy deposited in each of the slices. 
In calculations the beam energy spread was 
assumed to be Gaussian with FWHM of 0.2\% 

Secondary nucleons and nuclei can be created in $^{12}$C projectile fragmentation reactions at any point 
within the projectile range $R_p$, before the Bragg peak.  The range of a secondary fragment $R_{A,Z}$
with mass $A$ and charge $Z$ with the energy per nucleon $E$ equal to that of the projectile with mass 
$A_p$ and charge $Z_p$ can be expressed as $R_{A,Z}(E)=\frac{A}{Z^2}\frac{Z_p^2}{A_p}R_p(E)$, 
see e.g. Ref.~\cite{Fiedler:etal:2006}. Depending on the mass and charge of produced
fragments they either propagate beyond the Bragg peak, like e.g. protons, neutrons 
deutrons or helium nuclei, or stop before the distal edge of the peak, like e.g.
$^{10}$C or $^{11}$C projectile fragments. These properties of nuclear fragments were
already demonstrated by the calculations for carbon-ion beams within the MCHIT 
model~\cite{Pshenichnov:etal:2005,Pshenichnov:etal:2006}. We have found that also the recent 
data~\cite{Haettner:Iwase:Schardt:2006} are well described by the MCHIT model, including the
contribution of secondary particles beyond the Bragg peak, as shown in Fig.~\ref{fig:C12_ENERGY_TWO_PANEL}(a).

As one can see from  Fig.~\ref{fig:C12_ENERGY_TWO_PANEL}, the
fragmentation of projectile nuclei reduces the fluence of primary beam nuclei
progressively with increasing beam energy.
As found in Ref.~\cite{Haettner:Iwase:Schardt:2006}, at 400 A MeV beam energy about 
70\% of primary $^{12}$C ions change their charge due to
nuclear fragmentation reactions. Fortunately,  this does not deteriorate too much the
suitability of such beams for heavy-ion therapy. As seen in Fig.~\ref{fig:C12_ENERGY_TWO_PANEL}(a),
the peak-to-entrance ratio is still around 4, as compared to 6.5 at 200 A MeV.  

The carbon ions with energies above 500 A MeV are not suitable for
therapeutic use, as their ranges in human tissues extend beyond
typical dimensions of a patients' body. 
As follows from Fig.~\ref{fig:C12_ENERGY_TWO_PANEL}(b), at high energies the calculations predict 
a very large contribution of fragmentation reactions leading to the decreasing height
and disappearance of the Bragg peak. This makes the peak-to-entrance dose ratio unacceptably small.
As a result, the contribution from secondary nuclear fragments beyond the Bragg peak is very large.

This also demonstrates inefficiency of passive beam delivery systems 
in heavy-ion therapy. In this technique, the beam energy is adjusted by
inserting specially designed ridge filters, range shifters and compensators (boluses) in front of a patient. 
In calculations such beam modulating elements can be roughly represented by additional water
thickness. Secondary particles produced in beam fragmentation on such elements will be responsible 
for essential unwanted dose deposition beyond the Bragg peak.

\begin{figure}[htb]  
\begin{centering}
{\includegraphics[width=1.1\columnwidth]{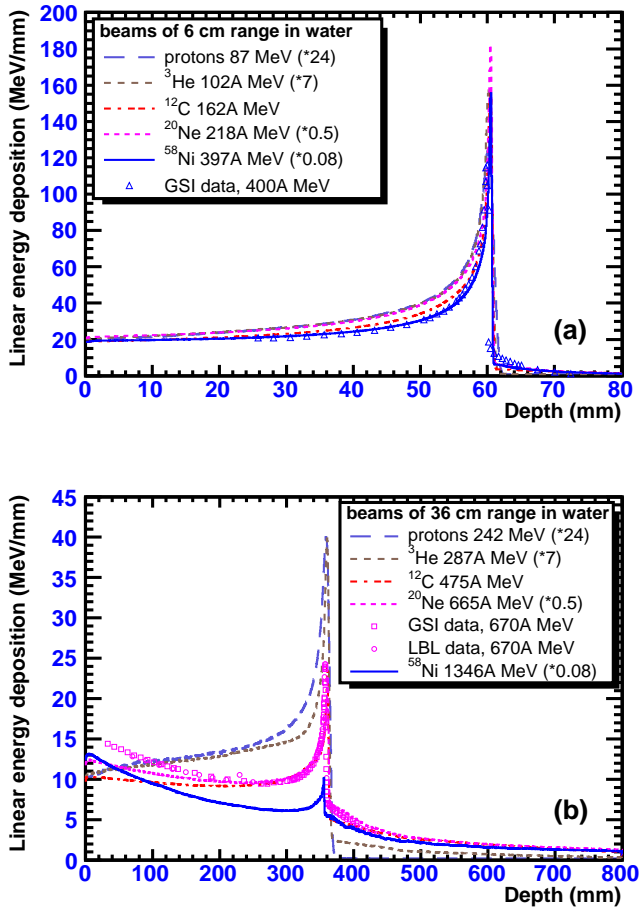}}
\caption{Color online. Calculated depth-dose distributions for (a) beams of protons (87 MeV), $^{3}$He (102A MeV),  
$^{12}$C (162A MeV), $^{20}$Ne (218A MeV) and $^{58}$Ni (397A MeV)   
nuclei and (b) for protons (242 MeV), $^{3}$He (287A MeV),  
$^{12}$C (475A MeV), $^{20}$Ne (665A MeV) and $^{58}$Ni (1346A MeV)  
nuclei in water (histograms). The distributions for all, but $^{12}$C nuclei, 
were multiplied by the rescaling factors (indicated in the legend) which normalize their entrance dose to
the one for $^{12}$C beam. Experimental data for $^{58}$Ni ions at 400 A MeV~\cite{Schardt:etal:1996}
and for $^{20}$Ne ions at 670 A MeV~\cite{Sihver:Schardt:Kanai:1998} 
are shown by various symbols.}
\label{fig:RANGE}
\end{centering}
\end{figure}

The depth-dose distributions for various ion species and beam energies are shown 
in Fig.~\ref{fig:RANGE} for two preselected positions 
of the Bragg peak at 60 and 360 mm, respectively. 
These distances represent two extreme cases of proton and heavy-ion therapy corresponding to 
shallow and deep sitting tumors.   
The beam energies for protons, $^{3}$H, $^{12}$C, $^{20}$Ne and $^{58}$Ne ions were chosen different 
in such a way that these projectiles have similar ranges in water, $\sim 60$ mm or $\sim 360$ mm.  
In Fig.~\ref{fig:RANGE}(a) the calculated distributions for all projectiles, but $^{12}$C,
were multiplied by the corresponding factors rescaling their entrance doses to the level of $^{12}$C ions. 
This facilitates the comparison of the depth-dose distributions for different beams.
Although the same factors were also applied in Fig.~\ref{fig:RANGE}(b),
this rescaling yields the same entrance doses only for protons and $^{3}$He.
For $^{20}$Ne and $^{58}$Ni projectiles the rescaled entrance dose is $\sim 20$\% higher than for $^{12}$C.     

Experimental data for $^{58}$Ni and $^{20}$Ne beams are also presented in  Fig.~\ref{fig:RANGE} 
for comparison. One can see that the agreement between the theory and the 
experiments is generally quite good.
We have tried to eliminate small discrepancies by fine tuning the beam energy  
for $^{58}$Ni and $^{20}$Ne within experimental uncertainties. 
We have found that calculations with 397 A MeV $^{58}$Ni beam energy can well describe the 
data of Ref.~\cite{Schardt:etal:1996} where 400 A MeV beam energy was quoted.  
Also, the Bragg peak position reported in Ref.~\cite{Sihver:Schardt:Kanai:1998}
for 670 A MeV $^{20}$Ne beam is better reproduced by calculations with beam energy of 665 A MeV.  
However, a noticeable discrepancy in shape between the MCHIT results 
and the data remains for $^{20}$Ne ions, as seen in Fig.~\ref{fig:RANGE}(b). 

Results presented in Fig.~\ref{fig:RANGE}(a) show that all beams from proton to Ni 
have a similar shape of the depth-dose distribution. In this situation the suitability of a specific beam for
the cancer treatment is dictated by its biological effects. In particular, the
$^{58}$Ni beam is probably not acceptable because of the very high entrance dose
(about 250 MeV/mm), which will destroy the healthy tissues.
By inspecting Fig.~\ref{fig:RANGE}(b) we conclude that
665 A MeV $^{20}$Ne and 1346 A MeV $^{58}$Ni beams are less suitable for cancer 
therapy compared to lighter projectiles due to increased beam
fragmentation. Indeed, the peak value for $^{20}$Ne is only twice as large as the entrance dose.
For $^{58}$Ni the situation is even worse: the Bragg peak is very weak and the entrance dose
is higher than the Bragg peak.  This shows clear advantage of lighter projectiles like protons, 
$^{3}$He and $^{12}$C ions for treatment of deep-seated tumors.


In summary, our systematic study of depth-dose distributions of nuclear beams in tissue-like medium
shows that the shape of the energy deposition profile depends rather on the ion range than on the ion mass.  
At relatively low beam energies corresponding to small penetration depths,
nuclear fragmentation reactions do not play a significant role, so that the Bragg peak is
well pronounced, and the tail due to secondary particles is small for all considered beams. 
In this case the depth-dose distributions for various nuclei are
similar to each other, and not very different from the distribution calculated for
carbon ions, which is currently the only modality used in ion therapy.
The peak-to-entrance dose ratio is large for all beam 
nuclei, from protons to nickel nuclei.

The depth-dose distributions for various nuclei are very different at high beam energies, i.e. for  
deeply penetrating beams. In this case one can benefit from using protons 
or $^{3}$He ions, with their largest peak-to-entrance dose ratio and minimal  
contribution from nuclear fragmentation. 
However, the Bragg peak for protons is less sharp compared to more heavy projectiles 
due to the longitudinal and lateral spread-out of the beam particles associated with the multiple 
scattering processes. We found that $^{3}$He beams can provide a reasonable alternative to
$^{12}$C beams in the case of large penetration depths, if the lateral beam scattering is not very crucial.
One needs experimental data on the depth-dose distributions of $^{3}$He beams to test the
validity of the MCHIT model for these ions.  

Here we have considered only distributions of physical doses from beams of various nuclei.
The proper assessment of suitability of a certain nuclear beam for cancer therapy should also include
its specific biological effects. As mentioned by several 
authors~\cite{Amaldi:Kraft:2005,Chen:Castro:Quivey:1981}, ions which are much heavier than carbon nuclei
have very high linear energy transfer even at the entrance point at patients' body that
makes difficult to save healthy tissues located at the beam path. As follows from our 
calculations, similar deficiency is observed already in the distributions of the physical doses for
nuclei heavier than $^{20}$Ne. Taking into account these observations we conclude that 
protons, $^{3}$He, $^{12}$C and, possibly, $^{16}$O are the most suitable beams  to be used in future
ion therapy facilities.

This work was partly supported by Siemens Medical Solutions.
We are grateful to Prof. Hermann Requardt for the discussions which stimulated the present study.
We are indebted to Dr. Dieter Schardt for useful discussions and for providing us the data tables for
the depth-dose distributions of carbon ions.

\end{document}